\documentclass[prb,twocolumn,superscriptaddress,showpacs,aps]{revtex4}

\usepackage{color}
\usepackage{graphicx}
\usepackage{bm}
\usepackage{amsmath}

\begin{document}

\newcommand{\Tc}{T_c}
\newcommand{\Hlo}{H_{c1}}
\newcommand{\Hup}{H_{c2}}
\newcommand{\Hupc}{H_{c2,c}}
\newcommand{\Hupab}{H_{c2,ab}}
\newcommand{\glam}{\gamma_{\lambda}}
\newcommand{\gH}{\gamma_H}
\newcommand{\xiab}{\xi_{ab}}
\newcommand{\xic}{\xi_c}
\newcommand{\lamn}{\lambda_n}
\newcommand{\dt}{\delta \theta}
\newcommand{\Happ}{\mathcal H}
\newcommand{\elam}{\varepsilon_\lambda}
\newcommand{\eH}{\varepsilon_H}
\newcommand{\Hparc}{\bm{\Happ} \parallel \bm{c}}

\title{Measuring the Penetration Depth Anisotropy in
       MgB$_2$ using Small-Angle Neutron Scattering }
% D.pal, L. DeBeer-Schmitt, T. Bera, R. Cubitt, C. D. Dewhurst, J. Jun, N. D. Zhigaldo, J. Karpinski, V. G. Kogan and M. R. %Eskildsen
\author{D. Pal}
\author{L. DeBeer-Schmitt}
\author{T. Bera}
\affiliation{Department of Physics, University of Notre Dame,
             Notre Dame, IN 46556}

\author{R. Cubitt}
\author{C. D. Dewhurst}
\affiliation{Institut Laue-Langevin, 6 Rue Jules Horowitz,
             F-38042 Grenoble, France}

\author{J. Jun}
\author{N. D. Zhigadlo}
\author{J. Karpinski}
\affiliation{Laboratory for Solid State Physics, ETH,
             CH-8093 Z\"{u}rich, Switzerland}

\author{V. G. Kogan}
\affiliation{Ames Laboratory and Department of Physics and Astronomy,
             Iowa State University, Ames, Iowa 50011}

\author{M. R. Eskildsen}
\email{eskildsen@nd.edu}
\affiliation{Department of Physics, University of Notre Dame,
             Notre Dame, IN 46556}

\date{\today}

\begin{abstract}
Using small-angle neutron scattering we have measured the misalignment between
an applied field of 4 kOe and the flux-line lattice in MgB$_2$, as the field
is rotated away from the $c$ axis by an angle $\theta$. The measurements,
performed at $4.9$ K, showed the vortices canting towards the $c$ axis for all
field orientations. Using a two-band/two-gap model to calculate the
magnetization we are able to fit our results yielding a penetration depth
anisotropy, $\glam = 1.1 \pm 0.1$.
\end{abstract}


\maketitle

%\section{Introduction}
%%%%%%%%%%%%%%%%%%%%%%%
The two-band/two-gap nature of superconductivity in magnesium diboride
(MgB$_2$) is well established.\cite{physicac} In this material supercarriers
reside on two distinctively different parts of the Fermi surface, with the 
larger energy gap ($\Delta_\sigma \approx 7$ meV) originating from $\sigma$
bonding of $p_{xy}$ boron orbitals, and the smaller gap ($\Delta_\pi =2.2$ meV)
arising from $\pi$ bonding of the $p_z$
orbitals.\cite{kortus01,liu01,choi02}
Furthermore the $\pi$ band is isotropic, while the $\sigma$ band is nearly
two-dimensional.

The anisotropy of a type-II superconductor is described either by
$\glam = \lambda_c/\lambda_{ab}$, where $\lambda$ is the magnetic penetration
depth, or by $\gH = \Hupab/\Hupc = \xiab/\xic$, with $\xi$ being the coherence
length.
Traditionally these two anisotropies have been considered to be identical, but
in materials with anisotropic gaps this is generally not the
case.\cite{kogan02a,kogan02b,miranovi03} Magnesium diboride represents an
extreme case where $\gH \neq \glam$. Theoretical work for this material
indicates $\glam(0) \approx 1.1$ and $\gH(0) \approx 6$ at low temperatures,
merging at $\glam(\Tc) = \gH(\Tc) \approx 2.6$ at the critical temperature
$\Tc$.\cite{kogan02a,miranovi03,golubov02}

While there is experimental consensus on the temperature dependence of
$\gH$,\cite{simon01,papavass01,angst02,budko02,zehetmay02,lyard04,rydh04} it
has recently been suggested that superconductivity in MgB$_2$ can be described
by a single anisotropy factor, $\gamma(H) = \glam(H) = \gamma_\xi(H)$, changing
from $\gamma(H=0) \sim 1$ to $\gamma(H=\Hup) \sim 6$. \cite{lyard04,angst04}
However, measurements of $\glam(H)$ are still contradictory, with flux-line
lattice imaging and torque magnetometry measurements yielding results ranging
from an arguably field-independent value of
$\glam \approx 1.2$,\cite{cubitt03a,eskildse03,fletcher05} to a strongly
field-dependent $\glam$ increasing with field and attaining a value of
$\sim 3.5$ at a modest field of 5 kOe.\cite{angst04,cubitt03b} Measurements
of the electronic specific heat in MgB$_2$, which mainly reflects $\gamma_\xi$,
yields an anisotropy which increases with the applied field.\cite{bouquet02}.

In this paper we will demonstrate an use of small-angle neutron scattering
(SANS) to determine the penetration depth anisotropy in MgB$_2$, by measuring
the misalignment between the applied field, $\bm{\Happ}$, and the direction of
the vortices in the flux-line lattice (FLL) as the field is rotated between the
$c$ axis and the basal plane.
% The misalignment is related to the torque
%$\bm{\tau} = \bm{M} \times \bm{\Happ}$ with the magnetization, $\bm{M}$,
%caused as the vortices to rotate away from the direction of the applied field.
Using a two-band/two-gap model we can fit the angular dependence of the
misalignment. At $4.9$ K and 4 kOe we find $\glam = 1.1 \pm 0.2$, smaller
than SANS measurements of the FLL anisotropy in single crystals\cite{cubitt03b}
and torque magnetometry\cite{angst04}, but in good agreement with SANS results
on powders\cite{cubitt03a} and scanning tunneling spectroscopy
results.\cite{eskildse03} The misalignment corresponds to vortices canting
towards the crystalline $c$ axis.

%\section{Experimental}
%%%%%%%%%%%%%%%%%%%%%%%
The experiments were performed at the NG3 SANS instrument at the NIST Center
for Neutron Research.
The sample was a $\sim 200 \; \mu$g platelike single crystal with the $c$ axis
parallel to the thin direction, grown using isotopically enriched $^{11}$B
to reduce neutron absorption\cite{karpinsk03}.
The $^{11}$B concentration in the sample is estimated to be $\sim 95$\%, with
$\sim 5$\% $^{10}$B due to the use of a natural boron BN crucible
(20\% $^{10}$B, 80\% $^{11}$B) in the crystal growth.
Incident neutrons with a wavelength $\lamn = 1$ nm and a wavelength spread
$\Delta \lamn/\lamn = 14$\% were used, and the FLL diffraction pattern was
collected by a 650 mm $\times$ 650 mm position sensitive detector with 5 mm
resolution. Both the collimation and sample-to-detector distances were 8 m. 
All measurements were performed in a magnetic field of 4 kOe applied at
different angles, $\theta = 0^{\circ} - 80^{\circ}$, with respect to the
crystalline $c$ axis. For all angles the field was rotated around the vertical
crystalline $[110]$ axis (see Fig. 1(a)), and the measurements were obtained
following a field cooling procedure to a temperature of $4.9$ K. The applied magnetic field is in the horizontal
direction and approximately parallel to the direction of the incident neutron beam.

%\section{Results}
%%%%%%%%%%%%%%%%%%%%%%%
\begin{figure*}
\includegraphics{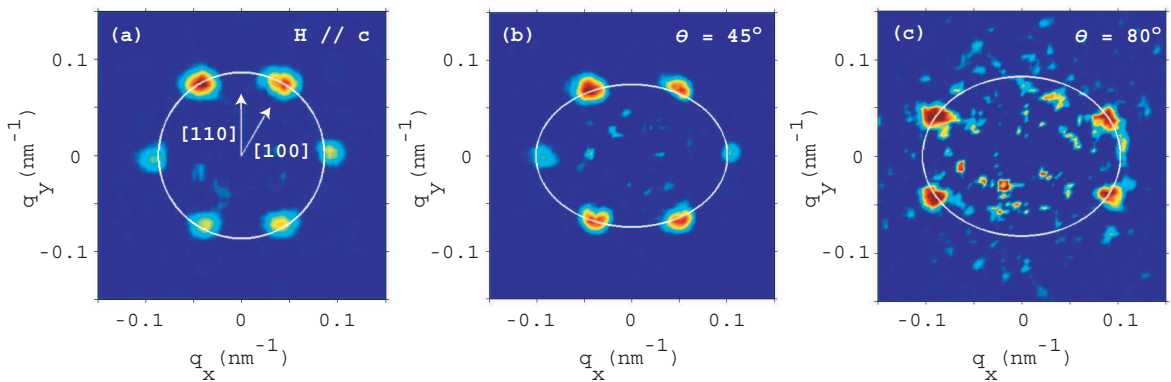}

\caption{(Color online)FLL diffraction patterns for MgB$_2$ obtained at 4 kOe and $4.9$ K,
         and with the applied field at angles $\theta = 0^{\circ}$ (a),
         $45^{\circ}$ (b), and $80^{\circ}$ (c) with respect to the $c$ axis.
         The orientation of the sample crystallographic axes is shown in
         panel (a). The reflections on the vertical axis in panel (c) are
         absent since they do not satisfy the Bragg condition. The data are
         smoothed by a $5 \times 5$ box average.
        }
\end{figure*}
Fig. 1 shows FLL diffraction patterns obtained for three different orientations
of the applied field. Each image is a sum of the scattering from the FLL as the
cryomagnet is rotated around the vertical axis in order to satisfy the Bragg
condition for the different reflections. 
In Fig. 1(a) the field is applied parallel to both the sample $c$ axis
($\theta = 0^{\circ}$) as well as the direction of the neutron beam.
Consistent with our earlier reports a hexagonal FLL is observed, with the
lattice plane normals (Bragg peaks) along the $a$ axis.\cite{cubitt03b}
Applying the field at an angle with respect to the $c$ axis leads to a
distortion of the hexagonal FLL as shown in Fig. 1(b), and eventually a
reorientation shown in Fig. 1(c). In contrast to $\Hparc$ where the
reorientation progresses continuouly above an onset field
$H^* \sim 5$ kOe,\cite{cubitt03b} the transition in rotated fields is first
order and occurs at $\theta \sim 70^{\circ}$ with an applied field of 4 kOe.
%Similar behaviour is observed in members of the borocarbide superconductors,
%where a continous symmetry transition observed with $\Hparc$ is replaced by a
%first order reorientation transition when the field is applied parallel to the
%basal plane.\cite{eskildse97,eskildse01} In the case of MgB$_2$
The FLL reorientation is attributed to the multigap nature of the
superconductivity in this material.\cite{cubitt03b,zhitomir04}
Previously the FLL anisotropy has been used as a measure of
$\glam$.\cite{cubitt03b,eskildse03} However, recent measurements have indicated
that in MgB$_2$ in the proximity of the reorientation transition 
this anisotropy may be influenced by vortex core effects,\cite{cubitt05}
similar to what is observed in members of the borocarbide
superconductors.\cite{eskildse01} In contrast to the FLL anisotropy, the direction of the
vortices in the FLL does not depend on vortex-vortex interactions in the plane perpendicular to
the flux lines. Hence measurements of the misalignment of the FLL with respect to the applied
field provides a method for determining $\glam$, which is not affected by vortex core effects.

The exact orientation of the applied magnetic field is obtained from the FLL
rocking curve with $\Hparc$. In this case the vortices are parallel to the
applied field.
In Fig. 2(a) the intensities of the Bragg peaks lying on the horizontal axis
are plotted as a function of the cryomagnet rotation angle, $\alpha$. The
calibrated zero ($\alpha = 0$) is  determined from the position of the midpoint
between the two peaks, obtained by Gaussian fits to the rocking curves.
\begin{figure}
\includegraphics{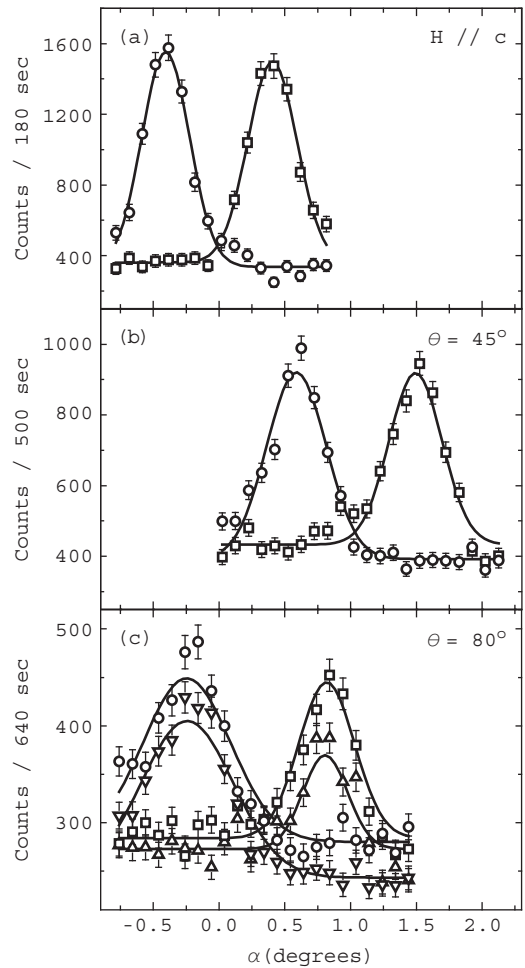}
\caption{FLL rocking curves for MgB$_2$ with the applied field at
         $\theta = 0^{\circ}$ (a), $45^{\circ}$ (b), and $80^{\circ}$ (c) with
         respect to the $c$ axis. The intensities at each angular setting,
         $\alpha$, is obtained by summing the detector counts at position of
         the Bragg reflections. The curves are Gaussian fits to the data. No
         background subtraction is performed.
        }
\end{figure}
As the field is rotated away from the $c$ axis the vortices in the FLL cant
away from the direction of the applied field as shown in Figs. 2(b) and 2(c).
The rocking curve midpoint now directly reflects the misalignment of the FLL,
$\dt$, with respect to the direction of the applied field. Note that above the
FLL reorientation transition four Bragg peaks shown in Fig. 1(c) are centered
around the horizontal axis and used to determine $\dt$. 
The misalignment corresponds to the vortices being rotated towards the
crystalline $c$ axis.
Only a slight increase in the average rocking curve width from
$0.45^{\circ} \pm 0.05^{\circ}$ FWHM ($\theta = 0^{\circ}$) to
$0.55^{\circ} \pm 0.05^{\circ}$ FWMH ($\theta = 80^{\circ}$) is observed as the
field is rotated away from the $c$ axis, indicating that no significant
disordering of the FLL takes place.
The experimental resolution is estimated to be $\sim 0.25^{\circ}$ FWHM. 
The absolute scattered intensity from the FLL decreases due to the increased
absorption, as the effective sample thickness becomes larger.

The results of the SANS measurements are summarized in Fig. 3. This shows a
symmetrical $\dt$ vs. $\theta$ dependence, with a maximum
$\dt \simeq 1^{\circ}$ close to $\theta = 45^{\circ}$, and decreasing towards
zero for $\theta = 90^{\circ}$ as expected. Measurements at angles above
$80^{\circ}$ were not possible due to absorption and increasing background
scattering from sample defects.
\begin{figure}
\includegraphics{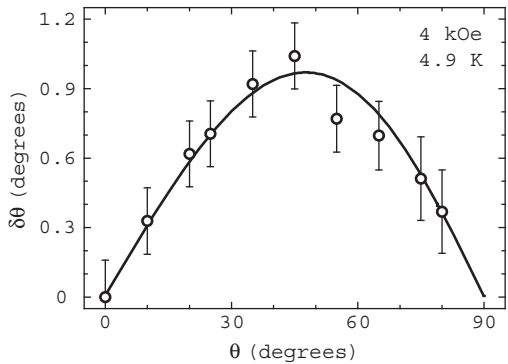}
\caption{FLL misalignment ($\dt$) as a function of the applied field rotation
         ($\theta$). The error bars take into account both the statistical and
         systematic errors of the measurements. The solid line is a fit to the
         data described in the text.
        }
\end{figure}

%\section{Discussion}
%%%%%%%%%%%%%%%%%%%%%%%
While the torque in a single-band superconductor is proportional to the
misalignment, $\dt$, this is not the case is materials with $\glam \neq \gH$
such as MgB$_2$. In the following it is assumed that the sample can be
approximated by a oblate spheroid (a coin) with the $c$ axis along $z$, and
choosing ($x,y$) so that the applied field $\bm{\Happ}$ lies in the $xz$-plane
forming an angle $\theta$ with $z$.
The linear relation between the applied field $\bm{\Happ}$, the induction
$\bm B$, and the magnetization $\bm M$ reads: \cite{LL}
\begin{equation}
   B_i = {\Happ}_i + 4\pi M_i (1-n_i) \,, \quad
   i = x,z \,,
   \label{B}
\end{equation}
where  $n_z=1-2n$ and $n_x=n$ are the demagnetization coefficients. For a
coin-shaped sample $n\ll 1$. The angle, $\theta_B = \theta - \dt$, between the
induction and the $c$ axis is given by
\begin{equation}
   \tan \theta_B
     = \frac{B_x}{B_z}
     = \frac{{\Happ}_x + 4\pi M_x (1-n_x)}
            {{\Happ}_z + 4\pi M_z (1-n_z)} \,.
   \label{tB}
\end{equation}
For $\Happ \gg \Hlo$ where $M\ll \cal H$, as is the case in the present
experiment, the misalignment $\dt = \theta - \theta_B$ is thus given by:
\begin{equation}
   \frac{\dt}{2\pi}
     = \sin 2\theta \left[ \frac{M_z}{\Happ_z} 2n -
                           \frac{M_x}{\Happ_x} (1-n) \right] \,.
   \label{dt}
\end{equation}

For materials with $\glam \neq \gH$ the free energy density is given
by:\cite{kogan02b}
\begin{equation}
   {\tilde F}
     = F - \frac{B^2}{8 \pi}
     = \frac{\phi_0 B \elam}{32 \pi^2 \lambda^2 \glam^{1/3}} \;
       \ln \frac{4\eta \Hupc \, \glam \, e}{B \elam \, (1+\beta)^2} \,,
   \label{F}
\end{equation}
where $\lambda = (\lambda_{ab}^2 \lambda_c)^{1/3}$ is the average penetration
depth,
$
%\begin{equation}
   \varepsilon^2_{\lambda,H} 
     = \sin^2 \theta_B + \gamma^2_{\lambda,H} \cos^2 \theta_B %\,,
%   \label{eps}
%\end{equation}
$,
and
\begin{equation}
   \beta
     = \frac{\glam}{\gH} \,
       \sqrt{\frac{B_x^2 + \gH^2 \, B_z^2}{B_x^2 + \glam^2 \, B_z^2}}
     = \frac{\glam \, \eH}{\gH \, \elam} \,.
   \label{beta}
\end{equation}
The constant $\eta$ lumps together the London uncertainty factor $\approx 1.4$
with $\pi\sqrt{3}/e$ yielding $\eta \approx 2.8$.

Evaluation of the magnetization,
$\bm{M} = - \partial {\tilde F}/\partial \bm{B}$, and the misalignment angle,
$\dt$, is now straightforward. Differentiating with respect to ($B_x,B_z$), and
noting that $B\elam = \sqrt{B_x^2 + \glam^2 B_z^2}$ one obtains $\bm{M}$ in
terms of $\bm{B}$:
\begin{widetext}
\begin{eqnarray}
   \frac{M_x}{M_0}
     & = & - \frac{B_x \glam^{-1/3}}{B \elam} 
             \left\{ \ln \frac{4\eta \Hupc \, \glam}{(1+\beta)^2 B \elam}
                     + \frac{2 \glam \, (\gH^2 - \glam^2 ) \cos^2 \theta}
                            {\gH \, (1+\beta) \, \elam \, \eH}
             \right\} \,,
   \label{Mx} \\
   \frac{M_z}{M_0}
     & = & - \frac{B_z \glam^{5/3}}{B \elam}
             \left\{ \ln \frac{4\eta \Hupc \, \glam}{(1+\beta)^2 B \elam}
                     - \frac{2 (\gH^2 - \glam^2) \sin^2 \theta}
                            {\glam \, \gH \, (1+\beta) \, \elam \, \eH }
             \right\} \,,
   \label{Mz}
\end{eqnarray}
\end{widetext}
where $M_0 = \phi_0/32 \pi^2 \lambda^2$.
Since $M \ll \Happ$, the angle $\dt$ is small and $M_i$'s in Eq. (\ref{dt}) can
be taken in the lowest approximation. Substituting Eqs. (\ref{Mx}) and
(\ref{Mz}) into Eq. (\ref{dt}) one can replace ($B, \theta_B$) with
($\Happ, \theta$) obtaining:
\begin{widetext}
\begin{equation}
   \frac{\dt}{2 \pi}
     = \frac{M_0 \sin 2\theta}{\Happ \glam^{1/3} \elam}
       \left\{ (1 - n - 2n \glam^2) \,
              \ln \frac{4 \eta \Hupc \, \glam}{(1+\beta)^2 \Happ \elam}
              + \frac{2 \glam \, (\gH^2 - \glam^2)}
                     {\gH \, (1+\beta) \, \elam \, \eH}
                \left[ (1-n) \cos^2 \theta + 2n \sin^2 \theta \right]
       \right\} \,.
   \label{dt2}
\end{equation}
\end{widetext}
For MgB$_2$, $\gH > \glam$ at all temperatures, and hence $\dt > 0$ for all
values of $\theta$ and $T$ provided that $n$ is small. Consequently, for a
coin-shaped sample, the vortices are expected to rotate towards the $c$ axis,
consistent with the experimental results.

The solid line in fig. 3 is the best fit of Eq. (\ref{dt2}) to the SANS data,
obtained by varying $\glam$ and $\lambda$, while setting $\gH = 6, \eta = 2.8$
and using an approximate demagnetization factor for the sample, $n = 0.01$.
This yields a value $\glam = 1.1 \pm 0.2$ and $\lambda = 155$ nm.
The results provides additional evidence for a low value
of $\glam$ even at this relatively high field. Due to the uncertainty on the
parameter $\eta$ as well as the demagnetization factor, $n$, a reliable
determination of $\lambda$ and $\gH$ is difficult, especially since changing
these two parameters both affects only the amplitude of $\dt$. Nonetheless,
the fitted value of $\lambda$ is within the range of literature values
summarized in Ref. [\onlinecite{golubov02}] and in fair agreement with
$\lambda = 136$ nm from Ref. [\onlinecite{angst04}]. However, we can not exclude a field dependent
$\gamma_{\xi}(H)$ merging with $\gH$ at $\Hup$, as suggested in Refs. [14, 16, 21]. 
On the other hand the profile of $\dt$ is determined solely by $\glam$,
allowing a reliable fit of this parameter. Increasing $\glam$ causes the
$\dt$-profile to become more asymmetric as shown in Fig. 4(a).
\begin{figure}
\includegraphics{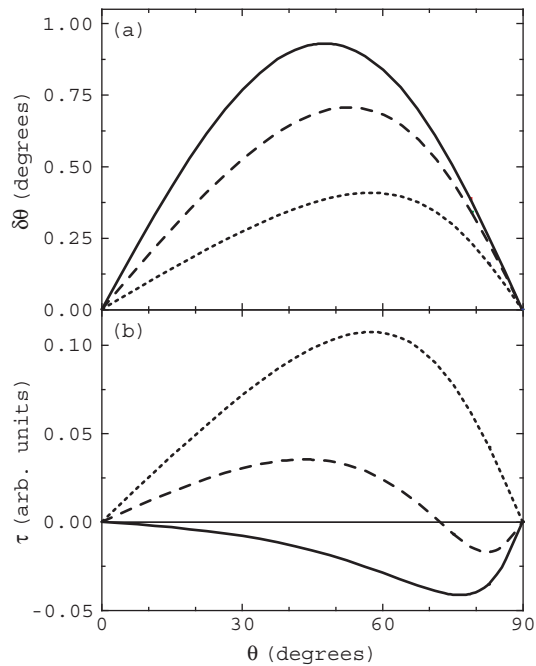}
\caption{Calculated misalignment (a) and torque (b) as a function of rotation
         angle, using Eqs. (\ref{dt2}) and (\ref{torque}) with $\Happ = 4$ kOe,
         $\lambda = 155$ nm and $n = 0.01$.
         The curves corresponds to low ($\gH = 6, \glam = 1.1$, solid line),
         intermediate ($\gH = 5.5, \glam = 1.5$, dashed line) and
         high ($\gH = \glam = 2$, dotted line) temperatures.\cite{miranovi03}
        }
\end{figure}
Although it would be possible to fit our data with $\gH = \glam = 1.1$ and
$\lambda = 135$ nm, this is an unreasonably low value of $\gH$ for a field of
4 kOe, and consequently the results presented here do not support a single
anisotropy factor as proposed in Refs. [\onlinecite{lyard04,angst04}].

Finally, it is of interest to compare $\dt$ to the torque density,
$\tau_y = M_z \Happ_x - M_x \Happ_z$. Inserting the $M_i$'s from
Eqs. (\ref{Mx}) and (\ref{Mz}) one obtains:\cite{kogan02b}
\begin{widetext}
\begin{equation}
   \tau_y
     = \frac{M_0 \Happ \sin 2\theta}{2 \glam^{1/3} \elam}
       \left\{ (\glam^2 - 1) \ln \left( \frac{4\eta \Hupc \, \glam}
                                             {(1 + \beta)^2 \Happ \elam}
                                 \right)
               - \frac{2 \glam \, (\gH^2 - \glam^2)}
                      {\gH \, (1 + \beta) \, \elam \, \eH} \right\} \,.
   \label{torque}
\end{equation}
\end{widetext}
Comparing this to Eq. (\ref{dt2}) it is clear that in general $\dt$ is not
proportional to $\tau_y$ as only the former is shape dependent. In particular
Eq. (\ref{torque}) suggests that at low temperature the torque is negative as
shown in Fig. 4(b), in contrast to what has been observed for layered
superconductors.\cite{tranquad88,farrell88} Near $\Tc$ one recovers
$\dt \propto \tau_y$ as in materials with $\glam = \gH$. Given the sucsess of
the model presented here in describing the misalignment, it would be of
interest to test further the relationship between $\dt$ and the torque.
Finally, one notes that for a spherical sample with
$n = 1/3$ one finds $\dt \propto \tau_y$ for all values of $\glam$ and $\gH$.

%\section{Summary}
%%%%%%%%%%%%%%%%%%%%%%%
In summary we have shown that SANS measurements of the misaligment angle
between the applied field and the FLL can provide information about the
penetration depth anisotropy in MgB$_2$. At $4.9$ K and 4 kOe we find a
relatively low value of $\glam = 1.1 \pm 0.2$. More studies are needed to
extend the measurements of $\glam$ to both  higher temperatures and fields, in
order to investigate the entire mixed phase of MgB$_2$.

\begin{acknowledgments}
%%%%%%%%%%%%%%%%%%%%%%%
We are grateful to J. Barker and P. Butler for their help during the SANS
experiment. We acknowledge the support of the National Institute of Standards
and Technology, U.S. Department of Commerce, in providing the neutron research
facilities used in this work. This work utilized facilities supported in part
by the National Science Foundation under Agreement No. DMR-0454672.
\end{acknowledgments}

% Bibliography
%%%%%%%%%%%%%%%%%

\newpage

\printfigures

\end{document}